\documentclass[12pt]{article}
\pdfoutput=1 
\usepackage{amsmath}
\usepackage{graphicx}
\usepackage[capitalise]{cleveref}
\crefformat{equation}{(#2#1#3)}
\usepackage{xcolor,soul}

\usepackage{url}
\usepackage{authblk}

\providecommand{\keywords}[1]
{
	\small	
	\textbf{\textit{Keywords---}} #1
}

\title{ \textbf{A quantum annealing approach to graph node embedding}}
\author{Hristo N.\ Djidjev}
\affil{Los Alamos National Laboratory\\ Los Alamos, NM 87545, USA \\
and \\
Institute of Information and Communication Technologies\\ Bulgarian Academy of Sciences, Sofia, Bulgaria 
} 

\date{}

\begin{document}
	
	\maketitle
	
	\begin{abstract}
		Node embedding is a key technique for representing graph nodes as vectors while preserving structural and relational properties, which enables machine learning tasks like feature extraction, clustering, and classification. While classical methods such as DeepWalk, node2vec, and graph convolutional networks learn node embeddings by capturing structural and relational patterns in graphs, they often require significant computational resources and struggle with scalability on large graphs. Quantum computing provides a promising alternative for graph-based learning by leveraging quantum effects and introducing novel optimization approaches. Variational quantum circuits and quantum kernel methods have been explored for embedding tasks, but their scalability remains limited due to the constraints of noisy intermediate-scale quantum (NISQ) hardware. In this paper, we investigate quantum annealing (QA) as an alternative approach that mitigates key challenges associated with quantum gate-based models. We propose several formulations of the node embedding problem as a quadratic unconstrained binary optimization (QUBO) instance, making it compatible with current quantum annealers such as those developed by D-Wave. We implement our algorithms on a D-Wave quantum annealer and evaluate their performance on graphs with up to 100 nodes and embedding dimensions of up to 5. Our findings indicate that QA is a viable approach for graph-based learning, providing a scalable and efficient alternative to previous quantum embedding techniques.
	\end{abstract}
	

	\section{Introduction}
	The vast amount of network data generated across various domains--ranging from protein interaction networks with thousands of molecules to social networks with billions of users--requires efficient methods for analyzing and extracting information from large graphs.
	One essential technique for transforming complex network data into a format suitable for machine learning applications is \textit{node embedding} \cite{cai2018comprehensive}. This method maps nodes to vectors such that similar nodes are represented by similar vectors, where the similarity metric depends on the specific application. The objective is to preserve structural and relational information while facilitating node comparisons, pattern detection, and the application of algorithms that do not directly operate on graphs. 
	
	Several methods using classical computing for node embedding exist such as DeepWalk \cite{perozzi2014deepwalk}, node2vec \cite{grover2016node2vec}, and GCNs \cite{kipf2016semi}. They have achieved strong performance and remain widely used in graph-based learning. However, classical computing approaches require a lot of computing power and may face challenges with scalability and computational efficiency, especially on large and complex graphs. This has led researchers to explore alternative computing paradigms.

	Quantum computing offers new possibilities to address some of these challenges by introducing novel methods for processing graph data. In particular, quantum machine learning (QML) methods \cite{schuld2021machine} provide new ways to process and analyze data, potentially improving expressiveness and computational efficiency across various machine learning tasks. Several approaches have been explored at the intersection of quantum computing and graph representation learning. Early works have investigated the use of quantum kernels for graph-based classification and clustering tasks \cite{schuld2019quantum, havlivcek2019supervised}. Variational quantum circuits (VQCs) have also gained attention as a way of learning node embeddings by parameterizing quantum states to encode graph structures \cite{benedetti2019parameterized}. These methods leverage the high-dimensional nature of quantum states for compact representation and efficient manipulation of graph data.
	
	One recent work relevant to this paper is the study of Zhou et al. \cite{zhou2024variational}, which proposes a variational quantum algorithm for node embedding. Their method constructs a parameterized quantum circuit that optimizes quantum states as node embeddings by minimizing a predefined loss function that aligns node similarities in the graph with the inner products of their embeddings in a Hilbert space. This approach encodes the graph’s adjacency structure while requiring only $O(\log N)$ qubits for a graph of $N$ nodes. The authors implement the proposed algorithm on a nuclear magnetic resonance quantum processor and show computational results.
	
	Despite the theoretical promise of quantum-enhanced embedding strategies, practical implementations remain limited by hardware constraints. Noisy intermediate-scale quantum (NISQ) devices struggle with limited qubit connectivity, susceptibility to noise, and training inefficiencies due to barren plateaus \cite{mcclean2018barren}. Furthermore, many QML algorithms assume the availability of a quantum database for rapid state preparation, an assumption some researchers have questioned \cite{aaronson2015read}. In the case of Zhou et al.’s approach \cite{zhou2024variational}, while it offers asymptotic efficiency, its practical performance is constrained by current NISQ technology. The implementation described in the paper is limited to graphs of four nodes and dimension two. Moreover, constructing an efficient circuit like the one described in the paper requires careful use of gates tailored to the specific structure of the input graph.
	
	In our work, we offer an alternative approach based on quantum annealing that is simpler and more scalable.
	
	Unlike quantum gate models, which rely on circuit-based operations and are capable of universal quantum computation, quantum annealing (QA) is a form of analog quantum computation specifically designed to solve combinatorial optimization problems by leveraging quantum mechanical effects such as superposition and quantum tunneling \cite{PhysRevE.58.5355, Morita_2008, das2008colloquium}. This approach is well-suited for graph optimization problems and has been demonstrated in tasks such as maximum clique detection \cite{chapuis2017finding}, graph coloring \cite{kwok2020graph}, and combinatorial optimization \cite{djidjev2018efficient}. QA operates by gradually transitioning a quantum system from an initial easy-to-prepare ground state to the ground state of a problem-specific Hamiltonian, following the framework of adiabatic quantum computation \cite{farhi2000quantum, RevModPhys.90.015002}.
	
	The practical implementation of QA is enabled by programmable quantum annealers such as those developed by D-Wave Systems, which currently offers processors with over 5000 qubits and 40,000 couplers \cite{dwave_advantage, king2022coherent}. These devices have demonstrated advantages in solving optimization problems, as well as other  applications, such as sampling from complex probability distributions \cite{nelson2022high} and quantum simulations of physical systems \cite{king2021scaling}. However, despite significant progress in both hardware and algorithms, QA has not yet been explored for solving graph node embedding problems.
	
	In this paper, we propose a  quantum annealing algorithm for the node embedding  problem. It formulates it as a quadratic unconstrained binary optimization  instance, which is a type of problem accepted by current quantum annealers. We implement the algorithm on a D-Wave quantum annealer and test its performance on graphs with up to 100 nodes and embedding dimensions of up to 5.
	
	The key contributions of this work include:	
	\begin{itemize}
		\item We show that current quantum annealers are well suited for solving node embedding  problems and may be useful for solving other problems in QML.
		\item Our algorithms are easy to implement and scale to larger inputs and dimensions compared to previous quantum methods.
		\item We  implement  our algorithms on a D-Wave quantum annealer and experimentally analyze their accuracy.
	\end{itemize}
	
		The remainder of this paper is organized as follows. \cref{sec:methods} describes the methods used in our approach. Section \cref{sec:results} presents the experimental setup and results. \cref{sec:discussion} discusses the findings and their implications. We conclude with discussion of the results and future research directions.
		
	\section{Methods}\label{sec:methods}
	In this section, we present our approach to node embedding using quantum annealing. We begin with background information on quantum annealing and QUBO models, followed by a formulation of the node embedding task as a QUBO optimization problem. We also discuss various strategies for handling constraints and different methods for defining similarities between pairs of nodes in the graph.
	
	\subsection{Quantum annealing and QUBO models}
		
	Quantum annealing (QA) is a metaheuristic approach for identifying the global minimum of an objective function by encoding it into a quantum system and leveraging quantum fluctuations to efficiently explore potential solutions. Unlike classical optimization, which often employs local search strategies, QA uses quantum tunneling to escape local minima and find high-quality solutions \cite{PhysRevE.58.5355, farhi2000quantum}.
	
	To solve an optimization problem using a D-Wave quantum annealer, it must be formulated as a minimization problem of the form
	\begin{equation}
		\text{minimize} \quad Q(\mathbf{x}) = \sum_{i<j} J_{ij} x_i x_j + \sum_i h_i x_i,
	\end{equation}
	where $\mathbf{x} = \{x_1, x_2, \dots, x_n\}$ represents a set of binary variables, while $J_{ij}$ and $h_i$ are real-valued coefficients encoding interactions and biases, respectively. When the variables take values in $\{-1,1\}$, the problem is referred to as an \textit{Ising model}. If the variables are restricted to $\{0,1\}$, the formulation corresponds to a \textit{quantum unconstrained binary optimization} (\textit{QUBO}) model. These representations are mathematically equivalent and can be converted into each other using a simple linear transformation.
	
	Minimizing a quadratic function over binary variables is an NP-hard problem \cite{Barahona1982}, while many important NP-hard problems  have simple QUBO or Ising formulations \cite{Lucas2014}. However, practical constraints such as limited connectivity, analog control errors, and qubit decoherence remain challenges in their widespread adoption \cite{D-Wave-ICE, pearson2019analog, king2022coherent}.
	Methods such as minor-embedding, quantum error correction,  and hybrid quantum-classical approaches help mitigate these limitations \cite{choi2011minor,pudenz2014error,djidjev2024enhancing}.
	
	Constructing a QUBO representation of an optimization problem typically involves two steps: first, the problem is expressed as a quadratic binary program with linear constraints, and second, those constraints are incorporated into the objective function to obtain an unconstrained formulation suitable for quantum annealing. The next section explores techniques for formulating QUBO representations of the node embedding problem.

	\subsection{The node embedding problem}
	
	In this section, we define the problem of learning node embeddings for a given $n$-node graph \( G = (V, E) \), where \( V \) is the set of nodes and \( E \) is the set of edges. Each node \( x \in V \) is assigned an unknown \( k \)-dimensional binary vector \( v(x) \in \{0,1\}^k \), which serves as its embedding in a high-dimensional space. The objective is to learn embeddings that preserve certain structural and relational properties of the graph.
	
	To formalize the definition, we introduce a \textit{similarity} function \( s: D \to [0,1] \), where \( s(x,y) \) represents the degree of similarity between nodes \( x \) and \( y \) and $D\subseteq V \times V $. A higher value of \( s(x,y) \) indicates greater similarity, while lower values suggest weaker relationships. The similarity function, which the user selects based on the application, can be derived from the graph structure, such as by considering adjacency relationships, weighted connections, or higher-order proximities.
			
	The goal is to learn embeddings such that the scaled dot product of two node embeddings, \( \frac{1}{k} ( v(x) \cdot v(y) ) \), approximates the given similarity score, i.e.,
	\begin{equation}
		\frac{1}{k} v(x) \cdot v(y) \approx s(x,y), \quad \forall x, y \in V.
	\end{equation}
	
	 To achieve this, we define an objective function that quantifies the total error across all pairs \( (x,y) \in D \) in approximating \( s(x,y) \) with \( v(x) \cdot v(y) \). The \textit{node embedding  (NE) problem}  is to find
	\begin{equation}
		\operatorname*{arg\,min}_{v \in \mathcal{V}} \sum_{(x,y) \in D} w_{xy} \Big(v(x) \cdot v(y) - s(x,y)\Big)^2, \label{eq:non_qubo}
	\end{equation}
	where $\mathcal{V} = \{ v: D \to \{0,1\}^k \}$ represents the set of all possible node embeddings and $w_{xy} \geq 0$ is a weight specifying the pair importance. To simplify formulas, we will set $w_{xy} =1$ for all pairs $(x,y)$, but the methods should work for arbitrary weights.

	Since the optimization function involves fourth-degree terms in binary variables, formulation \cref{eq:non_qubo} does not represent a QUBO problem. However, in the following sections, we propose methods to transform this formulation into a QUBO format that is acceptable by  a quantum annealer.
	
	\subsection{Formulating the NE problem as a QUBO}
	
	For each fixed pair of nodes $(x,y)$ and  \( v(x) =\{x_1,\dots,x_k\}\), \( v(y)=\{y_1,\dots,y_k\} \), its corresponding term in \cref{eq:non_qubo} 
	\[ \Big( \frac{1}{k} v(x) \cdot v(y) - s_{xy}\Big)^2= \left( \frac{1}{k} \sum_{i=1}^k x_i y_i - s_{xy} \right)^2 =
	\frac{1}{k^2} \left( \sum_{i=1}^k x_i y_i \right)^2 - \frac{2}{k}  s_{xy} \sum_{i=1}^k x_i y_i + s_{xy}^2,\]
	contains quartic terms and is therefore not a QUBO. To make the problem quadratic, we introduce auxiliary variables. 
	
		\subsubsection{Defining auxiliary variables}
		
		For each pair of nodes \( x \) and \( y \) and each dimension \( i \), we introduce an auxiliary binary variable \( z_{x_i, y_i} \) to represent the product of \(x_i\) and \(y_i\), i.e.,
		\begin{equation}
			z_{x_i, y_i} =x_iy_i. \label{eq:z_{x_i, y_i}}
		\end{equation}
		Using an additive squared penalty function $(z_{x_i, y_i} -x_iy_i)^2$ to enforce \cref{eq:z_{x_i, y_i}} will result in qubic terms.
		Hence, to enforce the equality using lower degree terms, we use a standard linearization technique for quadratic terms involving binary variables  \cite{Asghari2022Linearization}. It replaces the quadratic equality with a set of linear inequalities:
		\begin{equation}
			z_{x_i, y_i} \leq x_i , \label{eq:lin1}
		\end{equation}
		\begin{equation}
			z_{x_i, y_i} \leq y_i.  \label{eq:lin2}
		\end{equation}
		\begin{equation}
			z_{x_i, y_i} \geq x_i + y_i- 1.  \label{eq:lin3}
		\end{equation}

		While constraints \cref{eq:lin1}--\cref{eq:lin3} are linear, they cannot directly be implemented as squared penalty additive terms, since squared penalties work only for equality constraints. In order to enforce the inequalities, we define the following custom built penalty functions 
	\begin{align}
		P_1(x,y,i) &= z_{x_i, y_i} - x_i z_{x_i, y_i}, \label{eq:P1} \\
		P_2(x,y,i) &= z_{x_i, y_i} - y_i z_{x_i, y_i}, \label{eq:P2} \\
		P_3(x,y,i) &=  z_{x_i, y_i} - x_i z_{x_i, y_i} - y_i z_{x_i, y_i}+x_i y_i .  \label{eq:P3}
	\end{align}
	These penalty functions are designed to ensure that each inequality in \cref {eq:lin1}--\cref {eq:lin3} can be enforced using the corresponding function in \cref{eq:P1}--\cref{eq:P3}. Specifically, if we combine the three functions into $P(x,y,i)=\sum_{j=1}^3P_j(x,y,i)$, we claim that
	\begin{description}
		\item[(i)] if $z_{x_i, y_i} =x_iy_i$ then $P(x,y,i)=0$,
		\item[(ii)] if $z_{x_i, y_i} \neq x_iy_i$ then $P(x,y,i)\geq 1$. 
	\end{description}
This would imply that $P(x,y,i)$ can be used in a penalty term, e.g., by adding $\mu P(x,y,i)$ to the objective function with constant $\mu$ large enough. To prove the above claim, we just check that it holds for all possible binary assignment of $x=x_i,y=y_i,z=z_{xi,yi}$ as shown on \cref{tab:penalties}.
	
		\begin{table}[h!]
	\centering
	\begin{tabular}{|c|c|c||c|c|c|c||c|}
		\hline
		$x$ & $y$ & $z$ & $P_1$ & $P_2$ & $P_3$ & P & $z - xy$ \\
		\hline
		\textbf{0} & \textbf{0} & \textbf{0} & \textbf{0} & \textbf{0} & \textbf{0} & \textbf{0} &  \textbf{0} \\
		\hline
		0 & 0 & 1 & 1 & 1 & 1 & 3 & 1 \\
		\hline
		\textbf{0} & \textbf{1} & \textbf{0} & \textbf{0} & \textbf{0} & \textbf{0} & \textbf{0} & \textbf{0} \\
		\hline
		0 & 1 & 1 & 1 & 0 & 0 & 1 & 1 \\
		\hline
		\textbf{1} & \textbf{0} & \textbf{0} & \textbf{0} & \textbf{0} & \textbf{0} & \textbf{0} &\textbf{0} \\
		\hline
		1 & 0 & 1 & 0 & 1 & 0 & 1 & 1 \\
		\hline
		1 & 1 & 0 & 0 & 0 & 1 & 1 &-1 \\
		\hline
		\textbf{1} & \textbf{1} & \textbf{1} & \textbf{0} & \textbf{0} & \textbf{0} & \textbf{0} & \textbf{0} \\
		\hline
	\end{tabular}
	\caption{Values table for  functions $P_1=z - xz$, $P_2=z - yz$, $P_3=z - xz - yz + xy$, $P=P_1+P_2+P_3$, and $z - xy$. Bold values correspond to assignments where $z=xy$. One can see that if $z - xy=0$ then $P=0$ and if $z - xy\neq 0$ then $P\geq 1$.}\label{tab:penalties}
	\vspace{0.1cm}
\end{table}
			
		\subsubsection{Objective function}
		
		With the auxiliary variables \( z_{x_i, y_i} \) defined, we can approximate the scaled dot product \(\frac{1}{k} \,v(x) \cdot v(y) \) by summing over all those auxiliary variables. Let \( Z_{xy} \) represent this approximation, i.e.,
		\begin{equation}
			Z_{xy} = \frac{1}{k}\sum_{i=1}^k z_{x_i, y_i}.
		\end{equation}
		
		We aim for \( Z_{xy} \) to approximate \( s_{xy} \) for each node pair \( (x, y) \). The error for each pair can be estimated as \((Z_{xy} - s_{xy})^2\), leading to the following objective function for the NE problem:
		\begin{equation}
			\mathit{Obj}=\sum_{(x,y) \in  D }  (Z_{xy} - s_{xy})^2 = \sum_{(x,y) \in  D }  \Big(\frac{1}{k} \big( \sum_{i=1}^k z_{x_i, y_i}\big) - s_{xy}\Big)^2
			\label{eq:obj1}
		\end{equation}
		\begin{equation}
			= \sum_{(x,y) \in  D }  \left(  \Big( \frac{1}{k} \sum_{i=1}^k z_{x_i, y_i} \Big)^2 - \frac{2 s_{xy} }{k} \cdot \sum_{i=1}^k z_{x_i, y_i} + s_{xy}^2 \right).
			\label{eq:obj2}
		\end{equation}
		
		The first term, \(\Big( \frac{1}{k} \sum_{i=1}^k z_{x_i, y_i} \Big)^2\), further expands as:
		\begin{equation}
			\left( \frac{1}{k} \sum_{i=1}^k z_{x_i, y_i} \right)^2 = \frac{1}{k^2} \sum_{i=1}^k z_{x_i, y_i}^2 + \frac{2}{k^2} \sum_{i < j} z_{x_i, y_i} \cdot z_{xj, yj}. \label{eq:obj3}
		\end{equation}
		
		Plugging \cref{eq:obj3} into  \cref{eq:obj2} and ignoring the additive constant produces the final objective function in a QUBO format. Note that $ z_{x_i, y_i}^2= z_{x_i, y_i}$ for binary $ z_{x_i, y_i}$.
		
		\subsubsection{Adding constraints using the penalty method}
		
		To obtain the final QUBO, we combine the objective function with penalty terms using $P_1$, $P_2$, and $P_3$. This yields   function
		\begin{equation}
			Q= \mathit{Obj} + \mu \!\!\!\!\sum_{(x,y) \in D} \sum_{i=1}^k \sum_{j=1}^3 P_j(x,y,i) ,\label{eq:Q}
		\end{equation}
		where $\mathit{Obj}$ is defined in \cref{eq:obj1}--\cref{eq:obj3} and  \( \mu \) is a penalty parameter that controls the strength of the constraints on the auxiliary variables. Since $Obj$, $P_1$, $P_2$, and $P_3$ are quadratic binary functions, minimizing $Q$ is a QUBO problem.
				
		The penalty coefficient $\mu$ is usually determined experimentally. A small $\mu$ may lead to constraint violations, while an excessively large $\mu$ can make the optimization problem harder to solve numerically. In the next subsection, we will discuss setting penalty coefficients using the Augmented Lagrangian formulation.
		
	\subsubsection{Adding constraints using the augmented Lagrangian method}		
	
	The \textit{augmented Lagrangian method (ALM)} is an effective approach for constrained optimization that improves numerical stability and convergence by combining penalty terms with Lagrange multipliers \cite{Bertsekas1996}. ALM often outperforms quadratic penalty methods by reducing ill-conditioning and ensuring feasibility, but it relies on iterative updates.
	
	In general, for a constrained optimization problem of the form:
	\begin{equation}
		\min_{x} f(x) \quad \text{subject to} \quad g_i(x) = 0, \quad h_j(x) \leq 0,
	\end{equation}
	the augmented Lagrangian function is given by
	\begin{equation}
		\mathcal{L}(x, \lambda, \nu, \mu) = f(x) + \sum_i \lambda_i g_i(x) + \sum_j \nu_j h_j(x)
		+ \frac{\mu}{2} \Big(\sum_i g_i^2(x) + \sum_j \max(0, h_j(x))^2\Big),
	\end{equation}
	where $\lambda_i$ and \(\nu_j\)  are called  \textit{Lagrange  multipliers}.
	The  augmented Lagrangian function \(	\mathcal{L}(x, \lambda, \nu, \mu) \) is iteratively minimized and the values of \( \lambda_i \), \( \nu_j \), and \( \mu \) are updated  after each iteration by  the formulas
	\begin{align}
		\lambda_i^{(t+1)} &= \lambda_i^{(t)} + \mu^{(t)} g_i(x^{(t)}), \\
		\nu_j^{(t+1)} &=\nu_j^{(t)} + \mu^{(t)}  \max(0, h_j(x^{(t)})), \\
		\mu^{(t+1)} &= \rho \mu^{(t)},
	\end{align}
	where \( t \) denotes the iteration number, \(x^{(t)}\) denotes the solution from the previous iteration, and \( \rho > 1 \) is a scaling factor that controls the increase of \( \mu \). This iterative process continues until all constraints are satisfied or another stopping criterion is met.
	
	To derive a formula for the augmented Lagrangian version of the NE problem, we define constraint functions
	\begin{equation}
		C_1(x,y,i) = z_{x_i, y_i}  - x_i , \label{eq:lin12}
	\end{equation}
	\begin{equation}
		C_2(x,y,i) = z_{x_i, y_i}  - y_i , \label{eq:lin22}
	\end{equation}
	\begin{equation}
		C_3(x,y,i) = x_i + y_i - z_{x_i, y_i} - 1,  \label{eq:lin32}
	\end{equation}
	where we require \( C_j(x,y,i) \leq 0 \) for \( j \in \{1,3\} \) in any valid solution and all triples \( (x,y,i) \).
	
	Given these constraints, the augmented Lagrangian function for the NE problem can be defined as
	\begin{equation}
		\sum_{(x, y)} \Big( \left( Z_{xy} - s_{xy} \right)^2 
		+ \sum_{j=1}^3 \sum_{i=1}^k \big( \lambda_{j,i}C_j(x,y,i) + \frac{\mu}{2} \max\{0,C_j (x,y,i)\}^2\big) \Big),
	\end{equation}
	where \( \lambda_{j,i} \) (a shorthand for \( \lambda_{(x,y),j,i} \)) represents the Lagrange multipliers for each constraint \( C_j \) and dimension \( i \), and \( \mu \) is the penalty parameter adjusted iteratively along with all \( \lambda_{j,i} \).
	
	Instead of using in the penalty term the function \( \max\{0,C_j (x,y,i)\}^2 \), which is not quadratic, use its counterparts defined in \cref{eq:P1}--\cref{eq:P3}, which results in the augmented Lagrangian function
	\begin{equation}
		\mathcal{L} (\lambda, \mu) = \sum_{(x, y)} \Big( \left( Z_{xy} - s_{xy} \right)^2 
		+ \sum_{j=1}^3 \sum_{i=1}^k \left( \lambda_{j,i}C_j(x,y,i) + \frac{\mu}{2} P_j (x,y,i)\right) \Big). \label{eq:AL}
	\end{equation}
	     
	Plugging in the expressions for  $C_j$ and $P_j$ and simplifying, we get for the penalty terms, for each pair $(x,y)$
\[
\mathcal{L} (\lambda, \mu) (x,y) = \sum_{i=1}^k \Bigg( \Big(
\frac{\mu}{2} \left(z_{x_i, y_i} - x_i z_{x_i, y_i}\right) + 
\frac{\mu}{2} \left(z_{x_i, y_i} - y_i z_{x_i, y_i}\right) \]
\[ + \frac{\mu}{2} \left(x_i y_i + z_{x_i, y_i} - x_i z_{x_i, y_i} - y_i z_{x_i, y_i}  \right) \Big) \]
\[ + 
\Big(\lambda_{1,i} (-x_i + z_{x_i, y_i}) + 
 \lambda_{2,i}(-y_i + z_{x_i, y_i}) + 
 \lambda_{3,i}(-1 + x_i + y_i - z_{x_i, y_i}) \Big)
\Bigg)
\]
\[
= \sum_{i=1}^k \left((-\lambda_{1,i} + \lambda_{3,i})\, x_i +(-\lambda_{2,i} + \lambda_{3,i})\, y_i +  \left(\lambda_{1,i} + \lambda_{2,i} - \lambda_{3,i} + \frac{3\mu}{2} \right) z_{x_i, y_i} \right. \]
\begin{equation}
\left. + \frac{\mu}{2} \,x_i y_i - \mu\; x_i z_{x_i, y_i} - \mu \;y_i z_{x_i, y_i}  - \lambda_{3,i}\right), \label{eq:ALcoeff}
\end{equation}
which provides the coefficients to be used in the QUBO model  that corresponds to the pair $(x,y)$. 

\subsubsection{Augmented Lagrangian formulation with quadratic equality constraints}\label{sec:ALM}

In the previous versions, we linearized the constraint 
\begin{equation}
	z_{x_i, y_i} - x_i  y_i=0 \label{eq:eq}
\end{equation}
by using the linear constraints  \( C_j(x,y,i) \leq 0 \) for \( j \in \{1,3\} \) because the penalty term containing  \(C_j(x,y,i)^2\) is a quadratic function, while \( (z_{x_i, y_i} - x_i  y_i)^2\) is cubic. Hence, in the augmented Lagrangian function \cref{eq:AL}, we used $C_j$ for the  Lagrange multiplier terms and \(P_j\) for the quadratic penalty. But since $C_j$ are used to enforce \cref{eq:eq} and \(z_{x_i, y_i} - x_i  y_i\) is quadratic, i.e., consistent with a QUBO, we can use the latter instead of $C_j$ for the  Lagrange multiplier term while keeping $P_j$ for the quadratic one.

In order to derive the QUBO coefficients for this version, which we refer to as the \textit{augmented Lagrangian method with quadratic constraints (ALMQ)},
we define the equality constraint for each pair \((x, y)\) and each dimension \(i\) as:
\begin{equation}
	E(x,y,i) = z_{x_i, y_i} - x_i  y_i. \label{eq:equality_constraint}
\end{equation}
The new augmented Lagrangian function incorporates these constraints is 
\begin{equation}
	\mathcal{L} (\lambda, \mu) = \sum_{(x, y)} \Big( \left( Z_{xy} - s_{xy} \right)^2 + \sum_{i=1}^k \Big( \lambda_{i} E(x,y,i) + \frac{\mu}{2} \sum_{j=1}^3 P_j (x,y,i) \Big) \Big),\label{eq:ALeq}
\end{equation}
where \( \lambda_{i} \) are the Lagrange multipliers, \( \mu \) is the penalty parameter, and \( P_j \) are defined in \cref{eq:P1}--\cref{eq:P3}.

Using \(E(x,y,i)\) instead of \(C_j\) simplifies the formulation by reducing the number of constraints while still enforcing the necessary relationships. This results in a more compact QUBO representation, which may improve efficiency without negatively affecting solution quality.

Substituting \( E \) and \( P_j \) into the augmented Lagrangian function, the augmentation terms for each pair \((x, y)\) become
\begin{multline*}
	P(x,y)=\sum_{i=1}^k \Big( \lambda_{i} E(x,y,i) + \frac{\mu}{2} \big( P_1(x,y,i) + P_2(x,y,i) + P_3(x,y,i) \big) \Big)  \\
	=\sum_{i=1}^k \Big( \lambda_{i} (z_{x_i, y_i} - x_i \cdot y_i) + \frac{\mu}{2} \big( z_{x_i, y_i} - x_i z_{x_i, y_i} + z_{x_i, y_i} - y_i z_{x_i, y_i} + x_i y_i + z_{x_i, y_i} - x_i z_{x_i, y_i} - y_i z_{x_i, y_i} \big) \Big)\\
	=\sum_{i=1}^k \Big(\big( \lambda_{i}   + \frac{3\mu}{2}\, \big)\,z_{x_i, y_i}
	+\big(- \lambda_{i} +\frac{\mu}{2}\, \big)\,x_i y_i - \mu \,x_i z_{x_i, y_i} - \mu\, y_i z_{x_i, y_i}   \Big).
\end{multline*}

The coefficients $\lambda$ are updated using the formula
\[
\lambda_i^{(t+1)} = \lambda_i^{(t)} + \mu^{(t)}  \, E(x^{(t)} , y^{(t)} , i),
\]
where  \(t\) is the iteration number and \(E\) is defined in \cref{eq:equality_constraint}.

		\subsection{Defining similarity between nodes}\label{sec:similarities}
	In order to construct test instances for out methods, we need to assign a similarity score to each node pair. In our implementation, we consider three methods for computing similarity based on the Jaccard similarity and the  the graph structure.
	
	\subsubsection{Jaccard similarity for adjacent node pairs}\label{sec:jac}
	
	The \textit{Jaccard similarity }\cite{jaccard1901etude} between two sets \( A \) and \( B \) is defined as the size of their intersection divided by the size of their union:
	\begin{equation}
		J(A, B) = \frac{|A \cap B|}{|A \cup B|}\,.
	\end{equation}
	It quantifies the similarity between two sets based on their shared elements relative to their total distinct elements.
	
	Accordingly, we determine the similarity between two nodes \( x \) and \( y \) in a graph based on the similarity of their neighborhoods. We define the similarity score \( s_{xy} \) as
	\begin{equation}
		s_{xy} = J(N(x), N(y)) = \frac{|N(x) \cap N(y)|}{|N(x) \cup N(y)|},
	\end{equation}
	where \( N(x) \) and \( N(y) \) represent the sets of neighbors for nodes \( x \) and \( y \), respectively. To prevent cases where adjacent nodes can have zero similarity, we include the nodes themselves in their neighborhoods, i.e., \( N(x) \) is defined to include \( x \) for each node \( x \).

	We define the domain $D$ as the set of all node pairs for which $s_{xy} >0$. Note that a pair of nodes $(x,y)$ will have non-zero similarities exactly when their neighborhoods have a nonempty intersection, i.e.,  when the distance between them in graph $G$ is at most two. 
	
	\subsubsection{Including zero similarities pairs in the domain}		
	The previous similarity type uses only the non-zero similarities \( s_{xy} \), ignoring in the objective pairs with zero similarity. As a result, in the resulting embedding,  such pairs are not guaranteed to have small dot product. 
	In this subsection, we demonstrate that one can  include the zero similarities in the objective function without significantly increasing the hardness of the resulting QUBO. 
	
	For this purpose, we will add new terms to the objective function accounting for pairs with zero similarities. Let \((x,y)\) be such pair for which \(s_{xy}=0 \). We define the approximation error for \((x,y)\) as
	\[\mathit{Er}(x,y)=\left|\frac{1}{k}  v(x) \cdot v(y) - s_{xy}\right|  =\left|\frac{1}{k}  v(x) \cdot v(y)\right|=\frac{1}{k}  v(x) \cdot v(y), \]
	since $v(x)$ and $v(y)$ are binary vectors. Since \(\mathrm{Er}(x,y)\) is a polynomial of degree two, no auxiliary variables \( z_{xy} \) need to be defined, resulting in the following function for the zero-similarity pairs error:
	\[
	\mathit{Obj}_{\!0} = \sum_{(x, y) \in D_0} \mathit{Er}(x,y) =  \frac{1}{k} \sum_{(x, y) \in D_0} v(x) \cdot v(y) = \frac{1}{k} \sum_{(x, y) \in D_0} \sum_{i=1}^k x_i y_i,
	\]
	where \(D_0=\{(x,y)\in V\times V~|~s_{xy}=0\}\). Since there are no auxiliary variables, no additional penalty terms are needed to enforce constraints for this objective.
	
	The final QUBO that account for both zero and non-zero similarities can be obtained by adding \(\mathit{Obj}_{\!{0} }\) to  \cref{eq:Q}, \cref{eq:AL}, or \cref{eq:ALeq}, depending on which penalty method is being used.
		
	In some applications, it may be beneficial to weight the zero and non-zero similarity terms differently by using an objective function 
	\[
	Q_\mathrm{tot} = \alpha \mathit{Obj} + \beta \mathit{Obj}_{\!{0} },
	\]
	where \( \mathit{Obj} \) corresponds to pairs with non-zero similarities, and \( \alpha, \beta \) are weighting coefficients. Since these coefficients only scale existing terms, they do not affect the computational complexity of the problem.

	We also note that \( \mathit{Obj} \) measures squared errors (\( L_2 \) norm), while \( \mathit{Obj}_{\!{0}} \) uses absolute differences (\( L_1 \) norm). While this can be considered an inconsistency, it allows us to solve much larger problems on sparse graphs.

		\subsubsection{Graph adjacency similarity}
	
	In this metric, similarity between nodes \(x,y\), denoted by \( s_{xy} \), is determined directly from the graph adjacency matrix. Specifically, we define similarity as:
	\begin{equation}
		s_{xy} = A_{xy}\cdot J(\!N(x), \!N(y)),
	\end{equation}
	where \( A \) is the adjacency matrix of the graph, and \( A_{xy} \) is the entry indicating whether nodes \( x \) and \( y \) are directly connected (\( A_{xy} = 1 \)) or not (\( A_{xy} = 0 \)), and $J$ and $N$ are the Jaccard similarity  and node neighborhood functions, as defined in \cref{sec:jac}. 
	
	This similarity measure aims to learn direct connectivity in the graph. Nodes that share an edge are encouraged to have similar embeddings, while those without a direct connection are pushed apart. We use this similarity measure because it is similar to the one used in the implementation experiment in \cite{zhou2024variational} and because it allows us to apply the algorithms on larger instances, if the input graph is sparse. Compared to the Jaccard similarity from \cref{sec:jac}, where non-zero occurs when the nodes are at distance no more than two, in graph adjacency similarity that distance is just one. 
	
	\section{Results}\label{sec:results}
	
	\subsection{Experimental setup}
	We implemented our algorithms and ran experiments using the \texttt{Advantage\_system4.1} D-Wave  quantum annealer, which we accessed through the Leap quantum cloud service. 
	The quantum annealing parameters ($\texttt{num\_reads} = 1000$, $\texttt{annealing\_time} = 100$) were set based on typical values from prior studies to balance solution quality and computational efficiency. Unless stated otherwise, all other quantum annealing parameters remain at their default values.
	
	We generated random test graphs with numbers of vertices $n$ in \(\{10,15,20\}\) and average node degree $4$. We computed embedding with dimension $k\in \{2,3,4,5\}$ and  generated $s_{xy}$ similarities using the three metrics described in the previous section.  For each combination of parameters, we generated $5$ random graphs. 
	
	We compared six algorithms in total, varying by the optimization method  (quantum annealing or simulated annealing) and the constraint implementation method (squared penalty, ALM, ALMQ).  We selected simulated annealing for our experiments because, as a general-purpose heuristic for global optimization, it serves as a classical counterpart to quantum annealing. For  implementation, we used the simulated annealing algorithm provided in the Ocean D-Wave software. The ALM parameters were set to an initial $\mu$ of $0.5$, an initial $\lambda$ of $0$, and an increase factor $\rho$ of $1.1$. 
	
	We ran each (graph,similarity,algorithm) combination  \(3\) times.  The next subsection describe the results.

	\subsection{Error trends analysis }

	\subsubsection{Comparing the accuracies of the algorithms}
	
\begin{figure}
		\centering 
		\includegraphics[width=\textwidth]{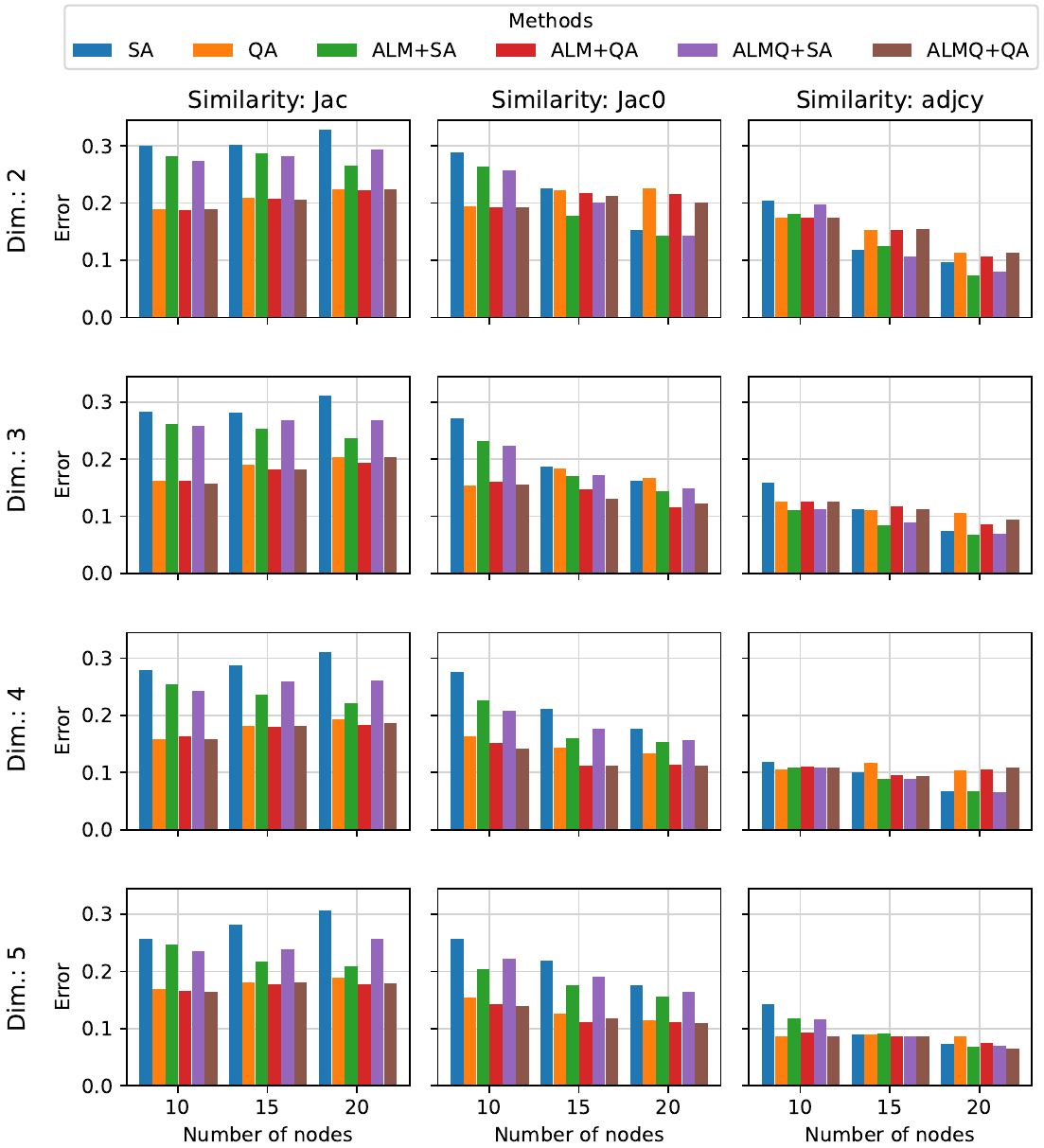} 
		\caption{Comparison of the methods for different classes of input graphs, embedding dimension, and similarity type.} 
		\label{fig:error_plots}
	\end{figure}

	To evaluate the effectiveness of the proposed algorithms, we analyze their error trends under varying experimental conditions, including the number of nodes, embedding dimension, similarity type, and algorithm. A list of the algorithms tested are given in \cref{tab:algs} and the three similarity types introduced in \cref{sec:similarities} are \texttt{Jac}, \texttt{Jac0}, and \texttt{adjcy}, respectively. The results are depicted in \cref{fig:error_plots}, where lower error values indicate better performance.
	
	\begin{table*}[!t]
		\renewcommand{\arraystretch}{1.5}

		\centering
		\begin{tabular}{|c|c|c|c|}
			\hline
			Name     & Constraints   & Optimization  & Classical/quantum \\ \hline\hline
			\texttt{SA}       & penalty & simulated annealing & classical         \\ \hline
			\texttt{QA}       & penalty & quantum annealing   & quantum           \\ \hline
			\texttt{AL+SA}    & ALM & simulated annealing & classical         \\ \hline
			\texttt{AL+QA}    & ALM & quantum annealing   & quantum           \\ \hline
			\texttt{AL+SA2}   & ALMQ           & simulated annealing & classical         \\ \hline
			\texttt{AL+QA2}   & ALMQ           & quantum annealing   & quantum           \\ \hline
		\end{tabular}
		\caption{Implemented algorithms used for the experimental analysis. ALM stands for the standard augmented Lagrangian method and ALMQ is for the augmented Lagrangian method with quadratic Lagrange multiplier term.}		
				\label{tab:algs}
	\end{table*}

	Comparing classical algorithms (SA-based) and quantum-assisted algorithms (QA-based), we observe a clear distinction. The quantum algorithms (QA, AL+QA, AL+QA2) generally achieve lower error rates than their classical counterparts (SA, AL+SA, AL+SA2), particularly for \texttt{Jac} and \texttt{Jac0} similarities. 
	
	However, there are a few exceptions. The plot for dimension 2, $n=20$, and \texttt{Jac0} similarity, as well as some cases with the \texttt{adjcy} similarity, show SA performing better than QA. 
	Explaining this deviation from the general trend is challenging since QA and SA are heuristics, and the conditions under which one algorithm outperforms the other are not fully understood.
	
	One possible explanation is that in very low-dimensional embeddings (such as dimension 2) and for simpler similarities like \texttt{adjcy}, the solution space is relatively easy to explore, reducing the advantage of quantum tunneling over classical optimization. Another factor could be that neither SA nor QA had their parameters optimized for performance; instead, default values were used to avoid bias. Notably, QA is highly sensitive to the embedding of the QUBO onto the quantum chip, a parameter that does not  affect SA.
	If QA and SA were tuned on a case-by-case basis, it is likely that QA would outperform SA in more, if not all, cases. However, fine-tuning these parameters would be labor-intensive given the large number of problem types and algorithms, and such analysis is beyond the scope of this paper.
		
	Comparing with respect to the constraint handling method, algorithms using the augmented Lagrangian approach (ALM and ALMQ) tend to achieve lower error rates compared to their penalty-based counterparts.
	On the other hand, when comparing ALM and ALMQ directly, their relative performance varies: in some cases, ALMQ has a lower error than ALM; in others, ALM performs better, while in many cases, their errors are roughly the same. The choice between the two can be made on a case-by-case basis, but ALMQ has the advantage of being simpler to implement than ALM.
	
	With respect to the embedding dimension, we observe a general trend across all algorithmic variants and graph sizes that increasing the  dimension tends to reduce the error for both classical and quantum algorithms. This trend holds consistently across all similarity measures but is more pronounced for \texttt{Jac0} and \texttt{adjcy}.
	
	The error trends with respect to the number of nodes vary depending on the similarity measure. For \texttt{Jac} similarity, as the number of nodes increases from 10 to 20, the error tends to increase slightly. However, for \texttt{Jac0} and \texttt{adjcy} similarities, we observe an opposite and more pronounced trend, where the error decreases as the number of nodes goes up. This may seem counter-intuitive, as larger problems are typically harder to solve accurately. However, this phenomenon can be attributed to the way similarity scores are computed. As the number of nodes increases while keeping the average degree fixed at 4, the ratio of zero to non-zero similarities also increases, because most node pairs become non-adjacent. This seems to simplify the problem, since zero similarities are easier to match.  However, for \texttt{Jac} similarity, where zeros are ignored, accuracy tends to decrease as the problem size grows.
		
	Finally, the three similarity types—\texttt{Jac}, \texttt{Jac0}, and \texttt{adjcy}—exhibit distinct behaviors. With \texttt{Jac} similarity, the error is highest overall, showing less variation across problem sizes and algorithms.
	Adjacency-based similarity tends to have lower error rates than \texttt{Jac0}, likely because it encodes fewer non-zero similarity pairs, resulting in a more structured optimization landscape. In contrast, \texttt{Jac0} assigns non-zero similarities to more node pairs, increasing complexity and introducing potentially conflicting requirements.
		
	\subsection{Contribution of zero similarities to the error}
	
	The Jaccard similarity metric (\texttt{Jac}) omits zero-similarity pairs from the objective function to simplify optimization, based on the assumption that these ignored pairs will naturally remain dissimilar in the learned embedding. To assess whether this assumption holds and to what extent, we analyze the effect of including zero-similarity pairs in the error estimation while still omitting them during optimization.
	
	We denote, as in the previous subsection, by \texttt{Jac} the method that ignores zero-similarities in the objective or error functions, and by \texttt{Jac0} the method where they are included. We compare three cases: (1) \texttt{Jac-Jac}, where both the objective and error estimation use \texttt{Jac}; (2) \texttt{Jac0-Jac0}, where both use \texttt{Jac0}; and (3) \texttt{Jac-Jac0}, where \texttt{Jac} is used in the objective but zero-similarity pairs are included in error estimation via \texttt{Jac0}. The difference between \texttt{Jac-Jac} and \texttt{Jac-Jac0} quantifies how much error increases when ignored zero-similarity pairs are accounted for in evaluation. Method \texttt{Jac0-Jac0} is included as a reference to explain possible deviations from expected error patterns. 
	
\begin{figure}
	\centering
	\includegraphics[width=1\textwidth]{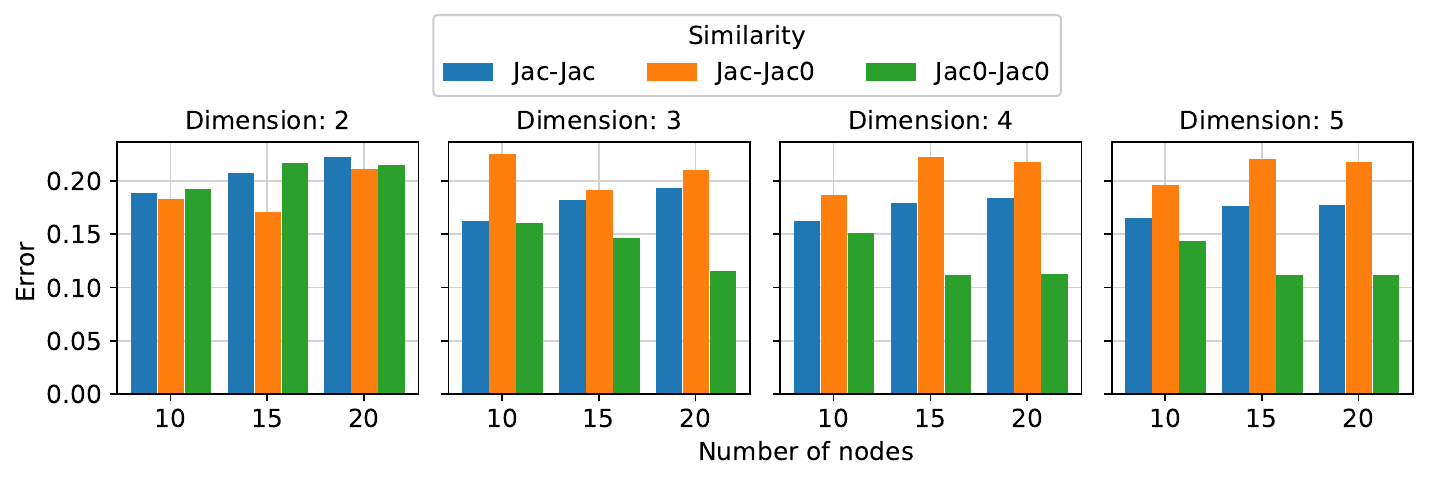}
	\caption{Error comparison across embedding dimensions for different similarity metrics. Labels indicate the similarity function used in the objective function (first component) and in the error estimation (second component).}
	\label{fig:zero_sim_error}
	\end{figure}

	Figure~\ref{fig:zero_sim_error} presents the error trends across different embedding dimensions as a function of the number of nodes. The results show that the \texttt{Jac-Jac0} error is higher than \texttt{Jac-Jac} for dimensions 3, 4, and 5, as expected, indicating that many of the ignored zero-similarity pairs do not remain dissimilar by chance.
	
	An exception occurs in dimension 2, where the \texttt{Jac-Jac} error is slightly lower than \texttt{Jac-Jac0} for $n=10$, as expected, but becomes higher for $n=15$ and $n=20$. A possible explanation for this discrepancy is that, as also observed in the previous subsection, the limited dimensionality of the solution space (two) makes it more difficult for QA to find accurate solutions.
	This also explains why \texttt{Jac0-Jac0}, which uses an error estimation function aligned with its objective, does not consistently outperform \texttt{Jac-Jac0}, where the error estimation and objective function are mismatched. The additional constraint imposed by zero-similarity pairs may distort the optimization landscape, increasing the difficulty for quantum annealers in two dimensions.
	For simulated annealing, as Figure~\ref{fig:error_plots} shows, this effect is not observed, and \texttt{Jac0-Jac0} achieves better accuracy than \texttt{Jac-Jac} for dimension two under SA.
	In higher dimensions (3, 4, and 5), Figure~\ref{fig:zero_sim_error} shows that \texttt{Jac0-Jac0} exhibits a lower error than the other methods, aligning with expectations.
	
	To summarize, these experiments show that in dimensions 3, 4, and 5, zero-similarity pairs shouldn't be ignored in the objective functions if they are considered important, while in dimension 2, more complex objective functions often produce distorted results.
				
	\subsection{Scalability} 
	\subsubsection{Dependence of the QUBO size on problem type}
	
	The size of a QUBO is a critical factor in determining whether it is solvable on quantum hardware, as larger instances may not fit on the quantum chip. Understanding how QUBO size scales with problem parameters is therefore important for assessing the feasibility of a given method. In our setting, QUBO size is influenced by the embedding dimension, the size and structure of the input graph, and the similarity metrics used.
	
\begin{figure}
		\centering
		\includegraphics[width=\textwidth]{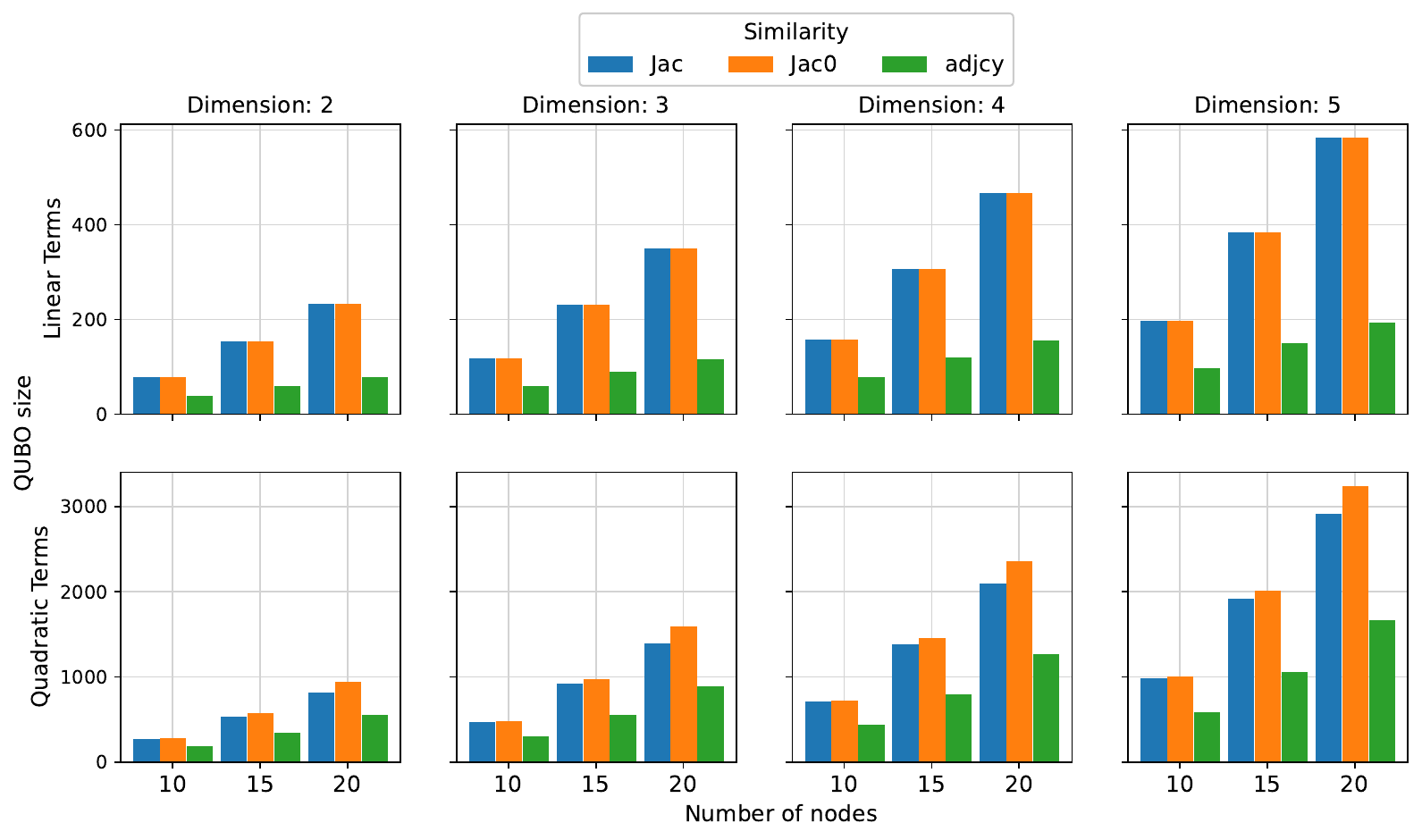}
		\caption{Dependence of QUBO size on the number of nodes, embedding dimension, and similarity metric. The top row displays the number of linear terms (number of variables), while the bottom row shows the number of quadratic terms.}
		\label{fig:facet_size}
	\end{figure}

	Figure~\ref{fig:facet_size} illustrates how these factors affect QUBO size. While the numbers of both linear and quadratic terms increase with the number of nodes, the rate of growth is more pronounced for quadratic terms, particularly in higher dimensions. Of the two parameters studied, embedding dimension has the stronger impact on QUBO size, while the number of nodes also has an effect, though to a lesser extent.
	
	Comparing QUBO sizes across the three similarity measures, we observe that the difference between the number of quadratic terms for the two Jaccard-based approaches,  \texttt{Jac0} and  \texttt{Jac}, is relatively small, and the numbers of the linear terms are exactly the same, indicating that the inclusion of zeroes in the Jaccard metric has a relatively small effect on QUBO size. In contrast, QUBO sizes resulting from  the \texttt{adjcy} similarity measure are consistently and substantially smaller than both \texttt{Jac0} and  \texttt{Jac} across all dimensions and node counts. 
	
	While QUBO size provides a measure of  a problem's complexity, the practical feasibility of solving these instances on a quantum annealer also depends on hardware topology constraints. Due to limited qubit connectivity in current quantum processors, logical QUBO variables often need to be mapped onto multiple physical qubits, increasing the effective problem size. In our experiments, this mapping was handled automatically by the D-Wave solver, but optimizing qubit allocation could further improve scalability and solution quality.
	
	In the next subsection, we analyze the maximum problem size that can be solved when the most scalable of the three studied metrics, \texttt{adjcy}, is used.
	
	\subsubsection{QUBO size scaling for graph adjacency similarity} 	
	
\begin{figure}
		\centering
		\includegraphics[width=0.99\textwidth]{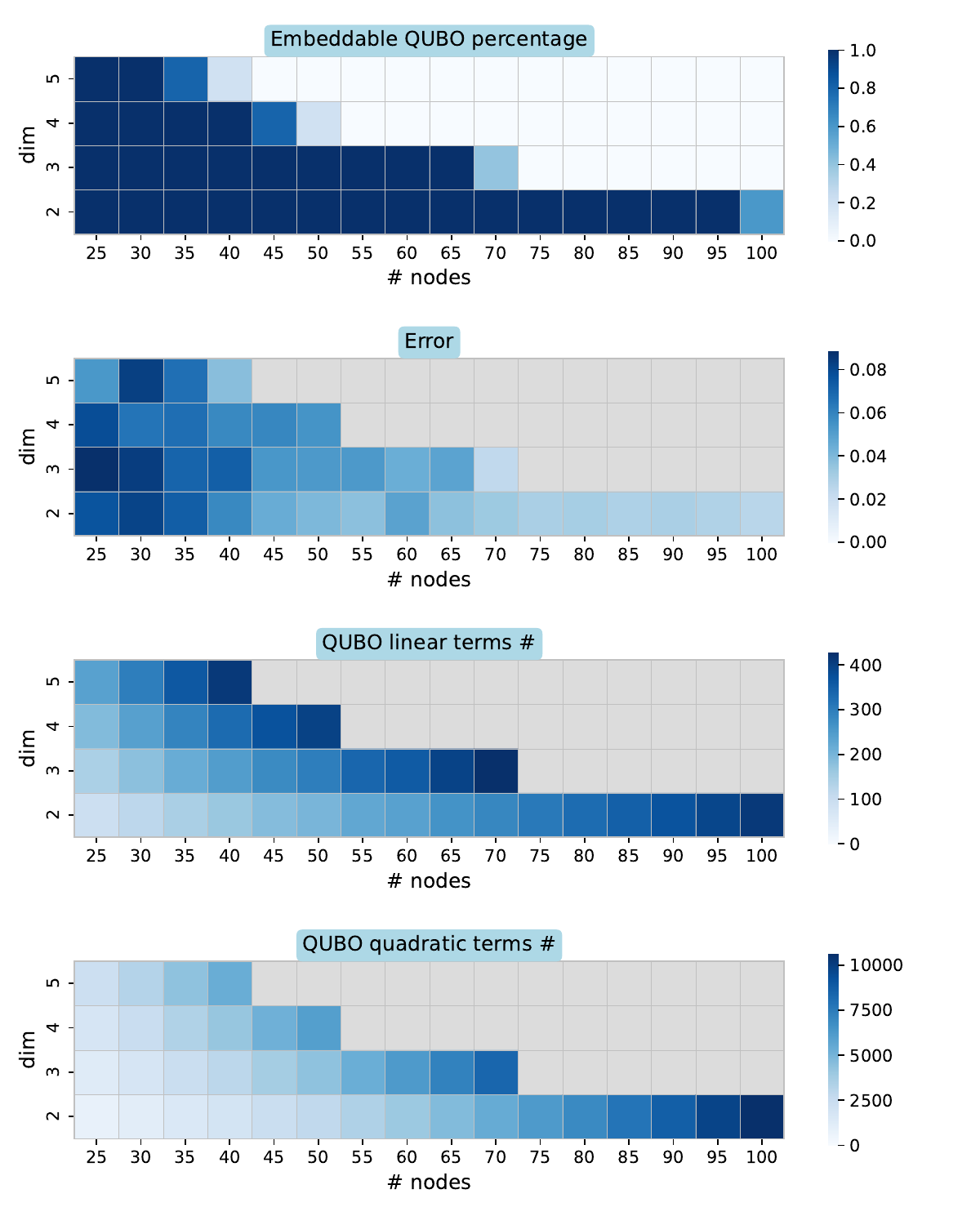}
		\caption{Scaling behavior of QUBO size and performance metrics for graph adjacency similarity. The first row shows the percentage of embeddable instances, the second row presents the QA error, and the third and fourth rows display the number of linear and quadratic terms, respectively.}
		\label{fig:adjcy0_sizes}
	\end{figure}

	To collect data, for each combination of problem size (number of nodes), between 25 and 100, and each embedding dimension, between 2 and 5, we generated five random node embedding problems and solved each of them five times on the quantum annealer.  
	\cref{fig:adjcy0_sizes} presents the results of our experiments.  
	
	The first row shows the percentage of instances where the generated QUBO could be embedded in the quantum annealer. As expected, embeddability decreases as the number of nodes and dimension increase, with higher-dimensional embeddings being especially difficult to embed. This indicates that embedding dimension has a stronger influence on feasibility than the number of nodes alone, confirming the trend observed in the previous subsection.  
	
	The second row of \cref{fig:adjcy0_sizes} displays the average error obtained using the penalty-based quantum annealing algorithm (QA). Errors range from 0.01 to 0.09, and these errors tend to decrease as problem size increases. While this might seem counterintuitive, it is a continuation of the trend we already observed in  \cref{fig:error_plots} for \texttt{adjcy} similarity, and the same explanation applies.
	
	The third and fourth rows show how QUBO's linear and quadratic terms, respectively, scale with the number of nodes and embedding dimension. This experiment is similar to the one in the previous subsection but extends to a larger range of node counts while focusing on adjacency-based similarity. The largest QUBO problems contain about 400 linear and 10,000 quadratic terms. 
	It is remarkable that such large QUBO problems are solvable on current D-Wave Advantage systems, which can embed dense QUBOs of up to 120 variables \cite{mcgeoch2020d}. This is particularly noteworthy given that the input graphs are randomly generated, i.e., without a special structure. A likely explanation is that the majority of quadratic terms are between local variables, e.g., \( x_i \), \( y_i \), and \( z_{x_i, y_i} \), making the problem of embedding the QUBO onto the quantum chip easier.  
	
	\section{Discussion}\label{sec:discussion}
	
	This paper explores the application of quantum annealing (QA) for graph-based node embeddings, introducing multiple problem formulations within the quadratic unconstrained binary optimization (QUBO) framework. The objective is to evaluate the feasibility and performance of QA-based node embeddings, tackling the same problem as prior quantum gate-model and variational quantum circuit study but with a fundamentally different approach.
	
	To enable compatibility with current quantum annealers such as D-Wave systems, we formulate the node embedding problem as a QUBO instance. We implement three similarity metrics—Jaccard similarity including only pairs with non-zero similarities (\texttt{Jac}), Jaccard similarity with zero similarities included (\texttt{Jac0}), and graph adjacency similarity (\texttt{adjcy}). We also implement and evaluate three different QA-based methods, distinguished by how they handle constraints: the standard approach using a quadratic penalty (QP) function and two approaches leveraging the augmented Lagrangian method—one incorporating linear Lagrange multiplier terms (ALM) and the other employing quadratic terms (ALMQ). Among these, ALMQ achieves the highest accuracy on average, followed closely by ALM, while the QP method exhibits a lower performance, suggesting that the augmented Lagrangian approaches lead to more accurate QUBO models. For each QA-based method, we also include its simulated annealing (SA)-based counterpart to compare QA against its classical analogue and assess performance differences. Our experiments on graphs of varying sizes demonstrate that QA outperforms SA on most problem types.
	
	The choice of similarity metric significantly impacts accuracy, with \texttt{adjcy} demonstrating the highest scalability. Using \texttt{adjcy} similarity, we solve larger problem instances for graphs with up to 100 nodes, a node degree of 4, and an embedding dimension of 5. In contrast, the previous variational quantum circuits (VQC) method \cite{zhou2024variational} was applied to graphs of 4 nodes and dimension 2.	
	
	The \texttt{Jac} similarity tends to yield  lower accuracy than \texttt{Jac0}  because it does not explicitly account for dissimilar node pairs. While \texttt{Jac} similarity provides a slight computational efficiency advantage compared to \texttt{Jac0}, its reduced accuracy limits its applicability.
	
	The results highlight the potential of QA as a scalable and efficient alternative for graph-based machine learning tasks. Compared to VQC, QA algorithms are simpler and more scalable for the current Noisy Intermediate-Scale Quantum (NISQ) hardware, although the VQC approach holds long-term potential for future quantum systems.
	
	Despite its potential, QA-based node embeddings face several challenges. The current hardware limitations, including restricted qubit connectivity, limited coherence times, and inherent noise in quantum annealers, impose constraints on the size and complexity of the problems that can be solved. Some of these issues can be mitigated by refining the implementation and optimizing tunable parameters. Another challenge is that QUBO problems inherently use binary decision variables, resulting in binary embeddings. However, we believe that a refinement of the proposed framework can also handle real-valued embedding vectors, which introduces a meaningful enhancement to the model. We plan to study this in future work. This enhancement would enable a more expressive embedding space and lead to improved optimization performance.
	
	\section{Acknowledgments}
	\label{sec:acknowledgments}
	This work was supported by the Laboratory Directed Research and Development program of Los Alamos National Laboratory under project 20210114ER. Los Alamos National Laboratory is operated by Triad National Security, LLC, for the National Nuclear Security Administration of U.S. Department of Energy (contract No.\ 89233218CNA000001).
	
	\bibliographystyle{plain}
	\bibliography{nodeEmbedding, references}

\end{document}